\documentclass[journal,onecolumn,12pt,draftcls]{IEEEtran}
\usepackage{graphics,graphicx,epsfig,amssymb,amstext,amsmath,algorithm,
algorithmic,wrapfig,multirow}
\usepackage{mathtools}
\usepackage{cite}
\usepackage{url}
\usepackage{times}
\usepackage{lipsum}
\usepackage{float}
\usepackage{placeins}
\usepackage{textcomp}
\usepackage[font=small,bf]{caption}
\usepackage{hyperref}
\usepackage{xcolor}
\usepackage{tabularx}
\usepackage{bm}
\usepackage{enumerate}
\usepackage{tikz}
\usepackage{pgfplots}
\usepackage{enumerate}
\usetikzlibrary{shapes,arrows,patterns}
\usepackage{subfig}
\usepackage{epstopdf}

\DeclareGraphicsExtensions{.eps}

\def\beq{\begin{equation}}
\def \eeq{\end{equation}}
\def\beqa{\begin{eqnarray} }
\def\eeqa{\end{eqnarray} }
\def\beqan{\begin{eqnarray*}}
\def\eeqan{\end{eqnarray*}}

\def\Exp{\mathbb{E}}

\def\tm1{t\! - \! 1}
\def\tp1{t\! + \! 1}

\def\UEs{\textsc{UE}}
\def\BSs{\textsc{BS}}
\def\GUE{G_{\textsc{UE}}}
\def\GBS{G_{\textsc{BS}}}
\def\gamDL{\gamma^{\textsc{DL}}}
\def\gamUL{\gamma^{\textsc{UL}}}
\def\TTTI{T_{\textsc{TTI}}}

\def\ULG{\textsc{ULG}}
\def\DLG{\textsc{DLG}}
\def\ULs{\textsc{UL}}

\def\ACK{\textsc{ACK}}
\def\ULACK{\textsc{ULACK}}
\def\DLACK{\textsc{DLACK}}

\newif\ifconf
\conffalse

\newif\ifonecol
\onecolfalse
\IEEEoverridecommandlockouts

\renewcommand{\footnoterule}{
  \kern -3pt
  \hrule width \columnwidth height 0.5pt
  \kern 3pt
}
\begin{document}
\tikzstyle{sig} = [draw=none, rectangle,pattern=north west lines, pattern color=blue!80, minimum height=4cm, minimum width=0.4cm]
\tikzstyle{sig2} = [draw=none, rectangle,pattern=north west lines, pattern color=red!80, minimum height=4cm, minimum width=0.2cm]

\title{Frame Structure Design and Analysis for
Millimeter Wave Cellular Systems}

\author{
    Sourjya Dutta,~\IEEEmembership{Student Member,~IEEE},
    Marco Mezzavilla,~\IEEEmembership{Member,~IEEE},
    Russell Ford,~\IEEEmembership{Student Member,~IEEE},
    Menglei Zhang,~\IEEEmembership{Student Member,~IEEE},
    Sundeep Rangan,~\IEEEmembership{Fellow,~IEEE},
    Michele Zorzi,~\IEEEmembership{Fellow,~IEEE}
    \thanks{This material is based upon work supported by the National Science
    Foundation under Grants No. 1116589 and 1237821 as well as generous support
    from NYU WIRELESS affiliate memberships.}
    \thanks{
        The authors are with NYU WIRELESS, New York University Tandon
        School of Engineering, Brooklyn, NY 11201 USA
        (e-mail: \{sdutta, mezzavilla, russell.ford, menglei, srangan\}@nyu.edu.  Michele Zorzi is with the University
        of Padova, (email: zorzi@dei.unipd.it)
    The conference version of this paper has been published in EuCNC 2016.}
 
}

% make the title area
\maketitle

\begin{abstract}
The millimeter-wave (mmWave) frequencies
have attracted considerable attention for fifth generation (5G) cellular communication as they offer orders of magnitude greater bandwidth than current cellular systems.
However, the medium access control (MAC) layer may need to be significantly redesigned to support the
highly directional transmissions, ultra-low latencies and high peak rates expected in mmWave communication. To address these challenges,
we present a novel mmWave MAC layer frame structure with a number of enhancements
including flexible, highly granular transmission times, dynamic control signal locations,
extended messaging and ability to efficiently multiplex directional control signals.
Analytic formulae are derived
for the utilization and control overhead as a function of
control periodicity, number of users, traffic statistics, signal-to-noise ratio and antenna gains.
Importantly, the analysis can incorporate
various front-end MIMO capability assumptions -- a critical feature of mmWave.
Under realistic system and traffic assumptions, the analysis reveals that the proposed flexible
frame structure design offers significant benefits over designs with fixed frame structures
similar to current 4G long-term evolution (LTE).  It is also shown that fully digital beamforming architectures offer significantly lower overhead compared to analog and hybrid beamforming under equivalent power budgets.
\end{abstract}

\section{Introduction}

The millimeter wave (mmWave) bands, roughly corresponding to
frequencies above 10~GHz,  have attracted considerable attention for
next-generation cellular wireless systems
~\cite{KhanPi:11-CommMag,rappaportmillimeter,RanRapE:14,andrews2014will,ghosh2014millimeter}.
These frequency bands offer orders of magnitude more spectrum than the
congested bands in conventional ultra high frequency (UHF) and microwave frequencies below 3~GHz.
In addition, advances in complementary metal-oxide semiconductor radio frequency (CMOS-RF) circuits and
the small wavelengths of mmWave frequencies enable large numbers of
electrically steerable antenna
elements to be placed in a picocellular access point or mobile
terminal providing further gains
via adaptive beamforming and spatial multiplexing.
Preliminary capacity estimates demonstrate that
this combination of massive bandwidth with large numbers of
spatial degrees of freedom can enable orders of magnitude
increases in capacity over current cellular systems
\cite{AkdenizCapacity:14,BaiHeath:14}.

However, the use of the mmWave bands for cellular is relatively new.
While mmWave systems have been successfully used for satellite
communications, cellular backhaul and wireless local area networks (LANs),
these applications consist generally of point-to-point
links with limited mobility \cite{EricssonBackhaul:13, Ted:60Gstate11, Baykas-WPAN:11, daniels200760, Daniels:10}.
Cellular systems require additional mechanisms to support
handover, track
channel conditions, and coordinate interference and traffic
between users, both within each cell and between neighboring cells.

In this work, we focus on one particularly important aspect of
this design problem -- namely the medium access control (MAC) layer \emph{frame structure}.
By the frame structure we mean the time-frequency
placement of all the relevant
MAC-layer channels, including the data, assignments, acknowledgements (ACKs),
and other control information to enable efficient use
of the spectrum resources by the cell.
While several groups have presented prototype designs
\cite{khan2012millimeter, pi2012millimeter, pi2014methods, kela2015novel, lahet2014achieving},
as we will see below, alternate design
choices can obtain dramatically improved overhead and utilization.

%The authors in \cite{andrews2014will} envision that 5G cellular systems will be
%highly heterogeneous and will require increased integration between different radio
%access technologies (RATs). Moreover, due to the propagation issues associated
%with mmWave links, it might also be necessary to concurrently use the 4G
%and 5G networks. Hence for our analysis,
%we assume a similar MAC layer structure as in 3GPP LTE~\cite{Dahlman:07}.
%Specifically, in the downlink, we assume that scheduling follows a similar
%hybrid automatic repeat request
%(H-ARQ) sequence as LTE with channel quality indicator (CQI) reports,
%downlink (DL) grants, DL data and uplink (UL)  ACK.
%Similarly, we assume the UL H-ARQ uses the LTE sequence of scheduling requests,
%UL grant, UL data and DL ACK.  The frame structure design problem
%is then how to allocate the various control and data channels
%to meet latency, overhead and other requirements.

\subsection{Design Requirements and Challenges}

The challenges in designing an efficient mmWave MAC-layer frame structure derive from the expectation that 5G cellular systems
will support extremely high peak data rates with very low latency~\cite{BocHLMP:14,ghosh2014millimeter}.
 Moreover, the radio infrastructure will be used by an increasing number of cellular users along with a high volume of machine-to-machine type communications \cite{shariatmadari2015machine}.
 In this work, we will consider the frame design under several different goals and constraints:
\begin{itemize}
\item \emph{Ultra low-latency:}  One of the most challenging goals is
the desire to obtain round-trip (base station (BS) to user equipment (UE) and back) airlink latencies of
approximately 1~ms.  Applications related to healthcare, logistics, automotive and
mission-critical control will require this stringent bound on the latency \cite{osseiran2014scenarios, shariatmadari2015machine}.
Moreover, as detailed in  \cite{fettweis20145g}, real-time cyber physical  experiences have similar
latency requirements.
This ultra low latency target is at least an order of magnitude
faster than the minimum latency currently offered by 3GPP LTE (10 ms, see \cite{network2011lte}).
Of course, the actual latency of the system will also depend on hardware processing
capabilities which are not within the scope of this study.  However, for our purpose,
we will consider a frame structure that can offer frequent opportunities for
transmission of data and control to meet these targets.

\item \emph{Multiple users:}  While 802.11 systems already
offer sub-millisecond latencies,
achieving similar low latency in a multi-user cellular system is significantly more
challenging.  Cellular systems depend on careful scheduling between multiple users
and cells
to efficiently use the airlink and achieve high levels of spatial reuse.
This scheduling demands significant control messaging.  As we will see below,
this control overhead grows with the number of users
and one of the main objectives of this paper is to find efficient ways to accommodate multiple
users and keep the ability to efficiently and rapidly allocate airlink resources.

\item \emph{Short bursty traffic:}  One of the main attractions of mmWave frequencies is the ability
to support multi-Gbps throughputs.
Cellular communication systems will need to efficiently
support these high data rates for both full buffer traffic and short bursty
transmissions.  Short transmission bursts may be needed for
radio resource control (RRC) layer messages, TCP ACKs, and applications that occasionally send short pieces of information.
Based on our analysis, we are the first to note that
the frame structure design has significant impact on the utilization of the airlink
in the presence of these short transmissions.  Additionally, in this work we show that frame
design not only depends on latency and overhead metrics, but also on the nature of the data traffic.

\item \emph{Beamforming architecture constraints:}
Due to the wide bandwidths and large number of antenna elements in the mmWave range,
it may not be possible from a power consumption perspective
for the mobile receiver to obtain high rate digital samples
from all antenna elements~\cite{KhanPi:11}.
Most designs have thus considered using either analog or hybrid beamforming
(at RF or IF) prior to the A/D conversion \cite{Heath:partialBF,SunRap:cm14},
or low-resolution fully digital front-end \cite{Madhow:ADC}
-- see a summary in~\cite{Heath2016mmWaveSP}.  The analog or hybrid beamforming architectures
can in effect ``look" in only one or a small number of directions at a time,
thereby placing severe restrictions on control channel multiplexing.
Conversely, low resolution fully digital architectures provide an energy-efficient way to multiplex simultaneously a large number of streams, but at the price of some signal quality degradation due to quantization errors.
\end{itemize}

\subsection{Proposed System Design}
To address these challenges, we propose a novel frame structure design that
incorporates several key changes relative to current LTE systems:
\begin{itemize}
\item \emph{Flexible TTI duration:} The current LTE system uses a fixed transmission
time interval (TTI) of one subframe (1~ms).
We show that this fixed TTI duration is extremely inefficient
when accommodating small packets and we propose a novel flexible TTI structure,
inspired by the design proposed in \cite{levanen2014radio}.

\item \emph{Directional control signaling:}  As discussed above, the mmWave
front-end architecture may limit the UE and/or BS ability to transmit and receive
in multiple directions at a time.  We propose and evaluate several alternatives
to multiplex control channels such as grants and channel quality indicator (CQI) reports,
under these constraints.  Moreover, this work is the first to provide designs that can exploit
low-resolution fully digital front-ends.

\item \emph{Extended control messages:}  One major contribution of this work is the use of extended messages for the uplink (UL)
and downlink (DL) grants as well as scheduling requests and capability reports. This is key for the use of dynamic TTI and flexible control signaling necessary for low-latency communication.
We show that these extended messages offer much faster scheduling with minimal increase in overall control overhead. 

\item \emph{Dynamic HARQ placement:}  In current LTE systems, the hybrid automatic
repeat request (HARQ) ACK is transmitted at a fixed time (3 subframes) after
the data transmission.  We are the first to propose that for mmWave systems, due to highly
variable packet sizes, decoding capabilities, and latency requirements, the HARQ ACK timing
should instead be scheduled dynamically.
\end{itemize}

Similar to the
LTE DL, we assume an orthogonal frequency division multiple access (OFDMA) waveform.
OFDMA has the benefit of simple equalization
and the ability to support orthogonal allocations in frequency and time.
Whether OFDMA is the optimal choice for mmWave cellular
remains to be determined.  However, our analysis
abstracts out the details of the particular waveform, and thus the concepts in our investigation can be applied to other
systems as well.

\subsection{Design Evaluation}  We present a novel
framework to evaluate two key performance criteria for the design options:
(i) control overhead and (ii) resource utilization for small packets.
The methodology is based on statistical distribution of signal to noise ratio (SNR) and
packet sizes and can thus be applied under a wide range of deployment
and traffic assumptions.   Our analysis also elucidates the
effect of different multi-input multi-output (MIMO) antenna architectures on MAC layer design.
In particular, we demonstrate the value of using low-resolution fully digital architectures.
The model is then used to assess various design options under realistic assumptions
for next-generation cellular evolution.  We demonstrate that
the proposed design can enable millisecond latencies with low control overhead while
accommodating large numbers of UEs in connected state.

\subsection{Organization}
Section \ref{sec:mimo} outlines the beamforming architectures available for mmWave MIMO antenna systems. In Section \ref{Sec:frameStruct} we discuss the possible frame structure design alternatives and propose a design for the control and data channels. A framework for the theoretical analysis of the MIMO architecture, frame structure and UL and DL channel design is presented in Section \ref{sec:dsgnana}. The results along with the realistic system assumptions are presented in Section \ref{sec:eval}. Section \ref{sec:conc} concludes the paper. Auxiliary technical details and accompanying discussions are given in the Appendix.

The conference version of this paper will appear in \cite{dutta2015Eucnc}.
This paper includes a more thorough and detailed design, particularly for all the control
signaling.  In addition, all the analytic derivations and details on the evaluation
methodology presented here were not included in the conference version.

\section{MIMO Architecture Models} \label{sec:mimo}
\subsection{Transceiver Architectures}

Before describing the channel structure, we need to consider
the different beamforming (BF) capabilities available at the base station.
As we will see, the MIMO processing assumptions
will have a significant impact on the control overhead and latency.

\begin{figure}
\centering
\subfloat[ ]{\includegraphics [page=1, trim={4cm 8cm 6cm 4cm},clip, scale=0.4]{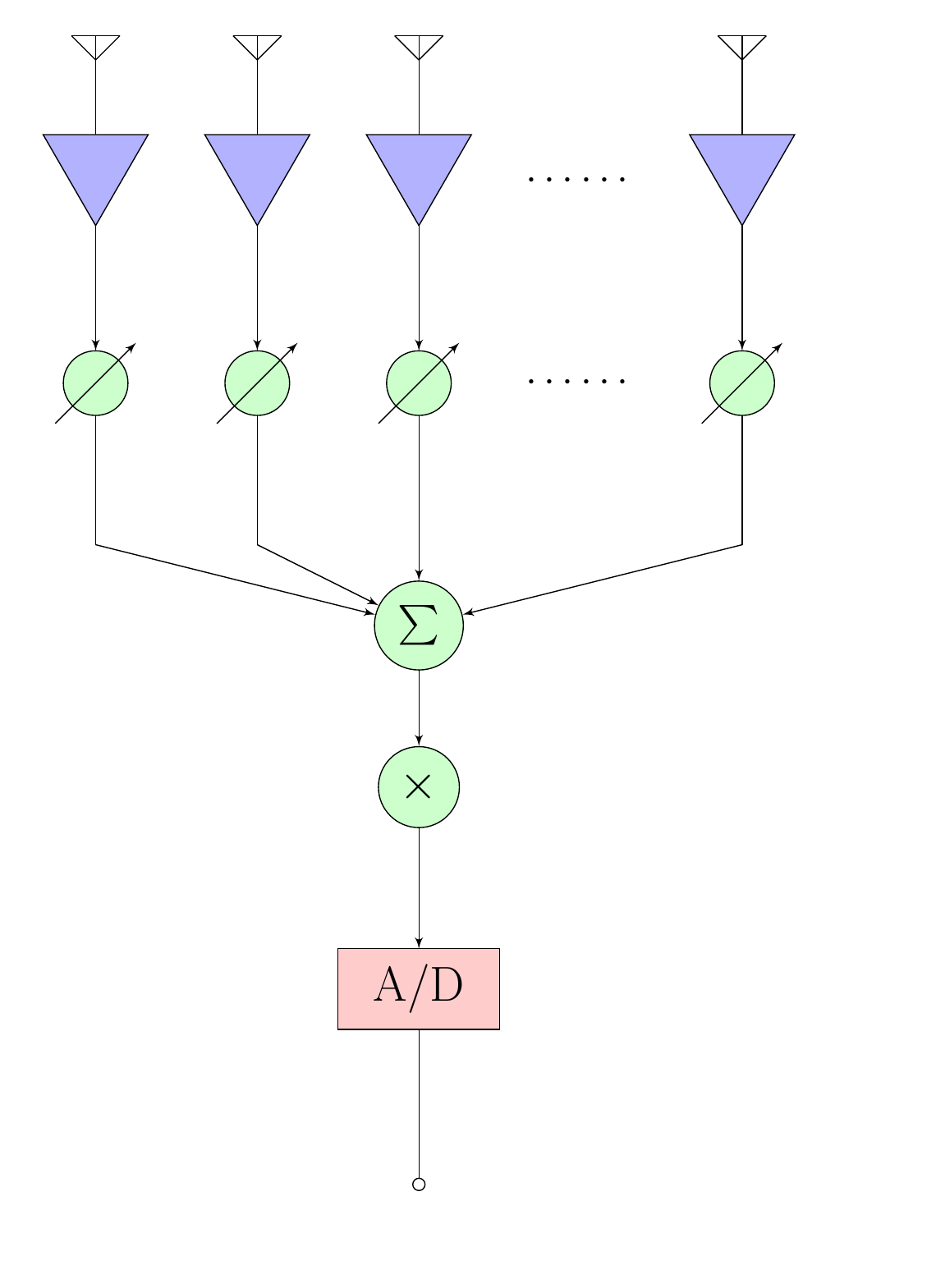} \label{BFAna}} 
\centering
\subfloat[ ]{\includegraphics [page=3, trim={4cm 8cm 6cm 4cm},clip, scale=0.4]{Beamforming.pdf} \label{BFDigi} }
\centering
\subfloat[ ]{\includegraphics [page=2, trim={4cm 8cm 3cm 4cm},clip, scale=0.4]{Beamforming.pdf} \label{BFHyb} }
\caption{Receiver architecture with (a) analog beamforming, (b) digital beamforming, and (c) hybrid beamforming.}
\label{Beamforming}
\end{figure}

In beamforming at conventional UHF and microwave frequencies,
there is typically a separate RF chain and A/D conversion
path for each antenna element.  This architecture enables the most flexibility
in that signals from the different antenna elements
can be combined digitally as shown in Fig. \ref{BFDigi}.
In the sequel, we will refer to this model as a \emph{fully digital}
architecture.

Unfortunately, in the mmWave range, due to the
large number of antenna elements and wide bandwidths,
it may not be possible from a power consumption perspective
for the BS receiver to obtain high rate digital samples
from all antenna elements~\cite{KhanPi:11}.
Most proposed designs thus
perform beamforming (at RF or IF) prior to the A/D conversion
\cite{KohReb:07,KohReb:09,GuanHaHa:04,Heath:partialBF,SunRap:cm14}.
This model saves power by using only one A/D or D/A but
the flexibility is reduced since the node can
beamform in one direction at a time.  This model, shown in Fig. \ref{BFAna},
is called \emph{analog beamforming}.

A hybrid between the digital and analog BF, shown in Fig. \ref{BFHyb}, is to have $K$
A/D or D/A conversion paths. Each of the $K$ paths corresponds
to a link between the BS and the UE called a stream.
This is called \emph{hybrid beamforming}
and was first proposed in \cite{Heath:partialBF}.  
In hybrid BF, the base station can beamform and combine to $K$ users
at a time with the full antenna gain. In the special case
when $K=1$, we obtain analog BF and when $K=N^{\rm ant}$,
the number of antenna elements, we get the capability of fully digital BF.

\subsection{Beamforming Gains}

To model the effect of directionality, let
$G_{BS}$ and $G_{UE}$ denote the maximum directional gain
achievable at the BS or UE when the beamforming is aligned
along the BS-UE link.  Due to reciprocity,
we assume the same gain is achievable in both transmit (TX)
and receive (RX) directions.
The exact directional gain will depend on the number of
antenna elements, multi-path angular scattering and
channel estimation accuracy, and may vary along different links.
However, to simplify the analysis, we assume that the maximum
gains are constant across all links and given by
\beq \label{eq:Gmax}
    G_{\rm BS} = N^{\rm ant}_{\rm BS}, \quad G_{\rm UE} = N^{\rm ant}_{\rm UE},
\eeq
where $N^{\rm ant}_{\rm BS}$ and $N^{\rm ant}_{\rm UE}$
are the number of antenna elements
available at the BS and UE.  The model in \eqref{eq:Gmax} holds exactly
 when the channel has a single angular path with no angular
dispersion~\cite{TseV:07}.  Studies of experimental data in \cite{AkdenizCapacity:14}
show that even in non line of sight (NLOS) channels with extensive
scattering and long-term beamforming we can generally obtain a gain
that is within 2~dB of this theoretical value, so we will make this
assumption in the paper.

For fully digital BF, we will assume that the maximum gains
can be obtained simultaneously
for an arbitrary number of UEs
since the BS can combine signals digitally.
In contrast, for analog BF, it can only obtain the maximum
directional gains for one user at a time.  For hybrid BF with $K$
streams, it can obtain the maximum gain\footnote{At the BS, both the
signal power and the noise power are split into $K$ parts,
 keeping the SNR constant. \cite{alkhateeb2014mimo}} for $K$ users at a time.
Note that, for hybrid BF, we have assumed that all antenna elements
are available to all streams via splitters or combiners.

With analog and hybrid BF, it may be necessary for the BS
to transmit to or receive from a number of users exceeding the
number of digital streams.  In this case, the BS will not
be able to obtain the full directional gain
and must set the antenna pattern in a wide angle to
transmit the signals to or receive them from
all the UEs.  We will let $G^{\rm omni}_{BS}$
denote the BS-side gain in this scenario.  A conservative lower
bound on this gain is given by
\beq \label{eq:Gomni}
    G^{\rm omni}_{\rm BS} = K,
\eeq
where $K$ is the number of digital streams ($K=1$ for analog BF).
In practice, the omni-directional gain may be larger if,
for example, the UEs are clustered angularly.  Also,
\cite{alkhateeb2014mimo} has considered the problem of optimizing
the $K$ streams to receive from multiple UEs with known spatial patterns.
This may increase the gain further.  However, in the analysis,
we will conservatively always assume an omni-directional
gain only as in \eqref{eq:Gomni}.

\subsection{Low Resolution Fully Digital BF}

One possible solution to the high power consumption of
fully digital architectures is to still have a full A/D conversion
path for each antenna element, but to use very few bits
per element -- say 2 to 3 bits per I/Q dimension.  Since
the power consumption of an A/D or D/A generally scales exponentially
in the number of bits (i.e., linearly in the number of levels),
using very low quantization resolutions can theoretically compensate
for the large numbers of parallel A/D and D/A paths.
This low resolution digital architecture has been considered in
\cite{Madhow:ADC,roufarshbaf2014ofdm,rajagopal2014power}. %-- see also \comment{Find other refs}.

To model the effect of low quantization resolution, we will use
an additive quantization noise model \cite{GershoG:92},
as used in \cite{FletcherRGR:07}.
As described in \cite{barati2015directional}, we can model
the effect of quantization as a reduction in the SNR.  Specifically,
if a channel is received with an SNR of $\gamma$ per antenna,
the SNR after quantization can be modeled as
\beq \label{eq:snrquant}
    \gamma' = \frac{\gamma}{1+\alpha \gamma},
\eeq
where $1/\alpha$ is an upper bound for the SNR, that depends on the quantizer design.

\section{Frame and Channel Structure} \label{Sec:frameStruct}

Having described the MIMO architectures, in this section we discuss in details the new flexible frame structure proposed in this work. The proposed UL and DL timelines are discussed to show how the control and the data channels need to be implemented to ensure low latency while properly utilizing radio resources. Additionally, we discuss the novel use of semi-static control signals and control signal multiplexing in the context of mmWave cellular systems.

\subsection{Flexible Frame Structure}
In current LTE systems -- see, for example, \cite{Dahlman:07} --
time is divided into regular intervals called \emph{subframes} of 1~ms,
with a fixed number of OFDM symbols per interval.
Each subframe is in turn divided into a control and a data portion.
The control portion is used for signaling various control messages such
as grants, ACKs, etc.   UL and DL transmissions occupy the entire
data portion of the subframe and hence the subframe duration is called
the transmission time interval (TTI).
In TDD LTE systems, the subframes are semi-statically
assigned as UL or DL.

Fixed TTI has been considered for mmWave cellular systems in \cite{KhanPi:11-CommMag}.
However, fixed TTI may lead to highly
underutilized subframes when transmitting very small MAC packet data units (PDUs).  In the current LTE standard,
this problem does not arise since small MAC PDUs can be allocated a very small portion
of the bandwidth (as small as one resource block),
with the remainder of the bandwidth being assigned to other UEs.
However, for a mmWave base station that can only direct its beam to one user
at a time, such FDMA scheduling is not possible, and thus the entire bandwidth must
be allocated for the entire subframe\footnote{While FDMA scheduling would still be
possible if the users are angularly close, this cannot be guaranteed in general.}.
The potential resource wastage is  particularly
dramatic for very wide bandwidth systems, as envisioned in the mmWave context.  We will
characterize this wastage precisely in Section~\ref{sec:smallpacket}.

In addition, the semi-static assignments of UL and DL subframes can lead to
poor utilization of the airlink when the traffic is asymmetric (i.e.,\ much higher in one direction).
In standard LTE TDD systems,
UL and DL directions cannot be assigned differently in neighboring cells since the DL transmissions
would overwhelm the uplink ones.  However,
mmWave links can be directionally isolated, and it is therefore useful to enable so-called dynamic
TDD scheduling where the UL and DL directions can be assigned dynamically
\cite{ghosh2014millimeter,garcia2015analysis}.
Thus we consider an alternate flexible frame structure, somewhat
similar to that proposed in \cite{levanen2014radio} but offering much greater flexibility for control signaling and scheduling.
Similar to LTE, we assume there is a larger frame period (possibly several milliseconds)
to reference periodic channels, but unlike \cite{levanen2014radio} there
are no fixed subframes within the frame.  Instead, control and data can be transmitted
in any OFDM symbol in either the UL or the DL direction.
TX-RX switching will be supported by guard periods.
Uplink symbol times are aligned at the BS.  Similar to LTE, this can be performed by
having the BS cell perform continuous feedback timing control so that the UEs
can advance their TX timing relative to their RX timing to ensure
that signals are properly aligned at the BS.

\begin{figure}[!t]
	\centering
	% Previously 0.53 textwidth
\subfloat[Downlink Timeline.]{\includegraphics [page=1,trim={1cm 17.5cm 0cm 0.75cm},clip, width=0.5\textwidth]{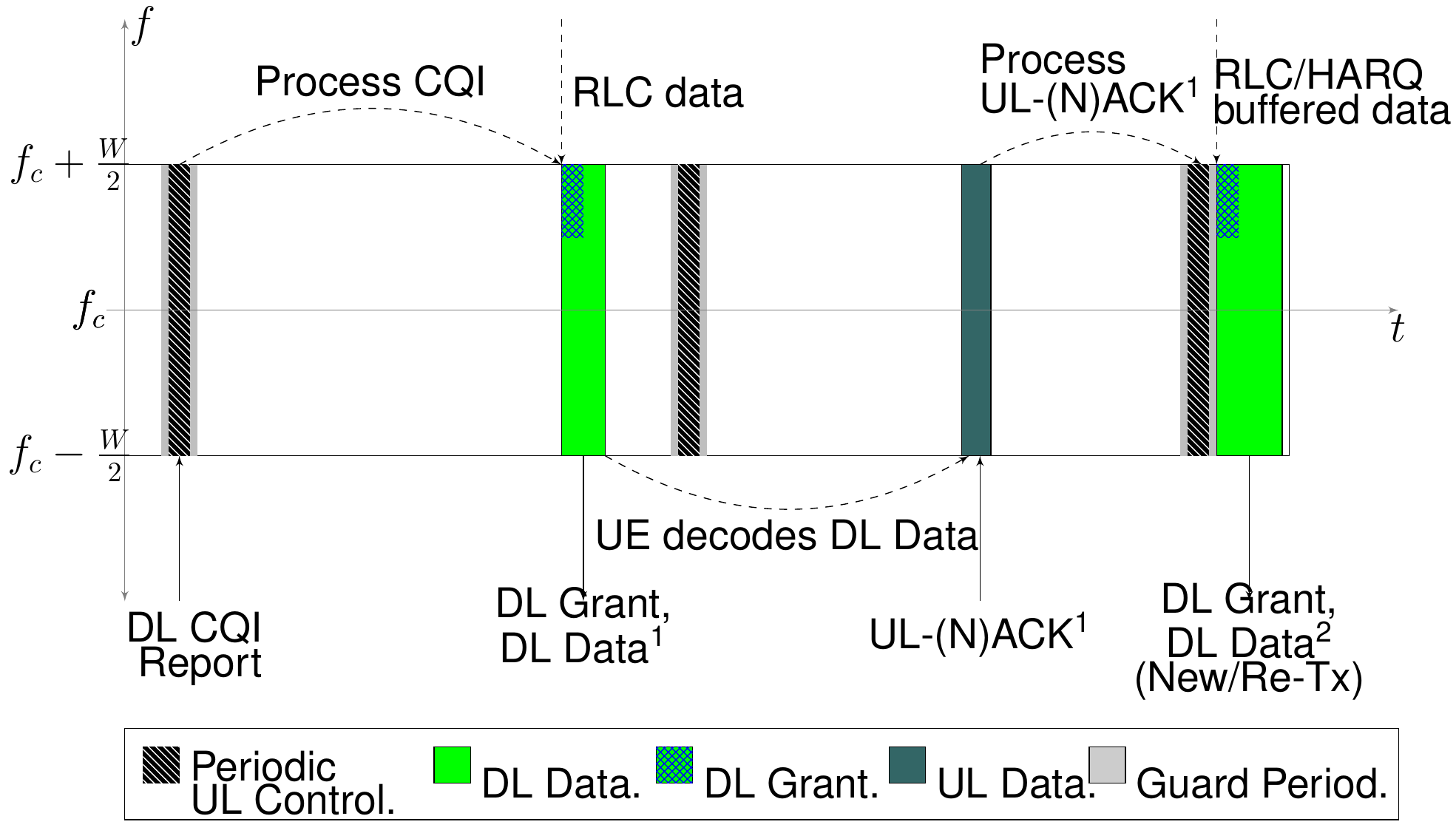}\label{fig:TDMAFlexa}}
\subfloat[Uplink Timeline.]{\includegraphics [page=2,trim={1cm 17.5cm 0cm 0.75cm},clip, width=0.5\textwidth]{TDMA_Flexible.pdf}\label{fig:TDMAFlexb}}
		\caption {Scheduling timeline for TDMA based mmWave systems with variable (i.e., dynamically scheduled) TTIs. $W$ represents the total bandwidth of the system. The center frequency is represented as $f_c$.}
		\label{fig:TDMAFlex}
\end{figure}

\subsection{Scheduling Timeline}

We illustrate the UL and DL scheduling using this flexible structure.

\paragraph*{DL Scheduling Timeline}
%We assume that the DL scheduling sequence follows the same basic steps as in LTE.
%To begin with,

The DL scheduling timeline is illustrated in Fig.~\ref{fig:TDMAFlexa}.
Each UE in connected mode is assigned dedicated uplink control
resources similar to the Physical Uplink Control Channel (PUCCH) in LTE.
These resources can be used to periodically transmit DL CQI
reports where the mobile periodically reports the SNR and other
channel characteristics for rate adaptation.
For a TDD system, the UL control channel can also be used for estimation
of the DL beamforming.  Note that the CQI reports will be continuously transmitted
for all UEs in connected mode, even when they are not scheduled to transmit data.

Now, suppose that DL radio link control (RLC) data arrives at the BS cell.  Since there are no fixed subframe boundaries,
the data can be scheduled at the next unassigned OFDM symbol, thereby enabling very low latency.
The assignment, called a DL grant, will be transmitted in
one of a fixed set of frequency locations in the
OFDM symbol and indicates the identity of the destination UE, the modulation and coding scheme (MCS) and other information to decode the data.  The UE scans each OFDM symbol for the
presence of this DL grant.  The data is transmitted along with DL grant.  This enables the data
and DL grant to obtain the full directional beamforming gain.

In the case that the BS supports analog beamforming with a single digital stream,
it can only transmit to one UE at a time.  Thus, the DL transmission (DL grant + data)
must occupy the entire bandwidth.  Hence, for short MAC PDUs, it may be necessary to transmit
for a very short TTI to avoid wasting the transmission time.  We thus propose
that the TTI duration is flexible and signaled within the DL grant using some additional information bits.  
%This will require
%some additional bits, that we will account for below.
Similar to LTE, we will assume that the system supports multi-process HARQ. The DL grant, hence, will also convey the HARQ process number, new data indicator and redundancy version. In case of retransmissions, the UE will know which prior transmissions to combine the current transmission with.  In addition, since the packet sizes and UE decoding capability may vary considerably in the mmWave space, we propose that the location of the UL ACK to be used
by the UE is also signaled with the DL grant.  Similar to the data, the UL ACK can be located
in any unassigned OFDM symbol.
A UE that has decoded the DL data successfully will transmit an UL ACK;
the absence of an ACK indicates a negative ACK due either to missing the DL grant
or failing to decode the DL data.

\paragraph*{UL scheduling timeline}    The UL scheduling timeline is shown in Fig. \ref{fig:TDMAFlexb}.  After a packet arrives at a UE, we assume that the UE sends a scheduling request (SR) to the BS, using the fixed UL control slots, to indicate that it has data to transmit.  Each UE must be allocated periodic resources to transmit the SR. Since we have access to a large number of degrees of freedom in the mmWave context, we will consider multibit SRs, resembling the LTE buffer status report (BSR). The multibit SR can enable a more fine grained report on the precise buffer levels at the UE and bypasses the additional overhead associated with using the one bit SR. The use of multibit SRs is especially important in the flexible TTI design as the amount of data the user requires to transmit determines the length of the transmission opportunity granted.

After decoding the SRs, the BS makes a scheduling decision that allocates OFDM symbols
to the UEs in the UL direction.  The UL grant can be transmitted in
 any OFDM symbol and shall contain the MCS, UL power control information and the
time location of the UL TTI and the DL ACK.
A UE that decodes the UL grant will then transmit the data in the TTI.
The time allocated for the UL grant, UL data and DL ACK can be dynamically adjusted
depending on the MAC PDU size as well as the UE processing capability for
preparing the transmission and BS processing capability for decoding the data.

\subsection{Semi-static Control Signals}
Many of the control signals such as scheduling requests
and CQI reports need to be allocated to dedicated resources whose
time locations would be fixed over
relatively long periods (e.g., reallocation requires a higher level RRC reconfiguration).
The dynamic UL and DL data transmissions would thus need to be allocated
``around" these
semi-statically fixed transmissions.  One simple mechanism would be to assign
UL and DL grants in sets of discontinuous time intervals,
thereby allowing data TTIs to ``skip" over any pre-assigned control signals.
We will account for this extra signaling overhead in Section~\ref{sec:overhead}.

\subsection{Control Signal Multiplexing} \label{sec:digBF}

As discussed in Section \ref{sec:mimo}, a BS that has analog BF with only
a single digital stream, can transmit or receive in one direction at a time.  On the other hand,
BS cells with hybrid or low-resolution
fully digital beamforming may transmit or
receive in multiple directions simultaneously.  When communication in
multiple simultaneous directions is possible,
control channels for different UEs can be multiplexed together in the same OFDM symbol
at different frequencies.  We will see in our analysis below that, since the control
messages are short, this frequency division multiplexing (FDMA) enables significantly
lower overheads.

\section{Design Analysis}\label{sec:dsgnana}

In this section we develop a general framework for evaluating two key performance metrics
for the frame structure design:  overhead and utilization.  We develop a general methodology that can
incorporate different assumptions on the statistical distributions of SNRs
and traffic patterns.  Importantly, the evaluation will enable us to quantify the
benefits of fully digital transceivers.  We will also be able to
compare the proposed flexible TTI structure to what we will call a fixed TTI
system where the TTIs occur at fixed boundaries with fixed durations.
We detail the evaluation model in Subsection \ref{subsec4a}. This is followed by the derivation of the utilization factor in Subsection \ref{sec:smallpacket}. In Subsection \ref{sec:overhead} we derive analytical expressions for the overhead of the control messages.

\subsection{Evaluation Model} \label{subsec4a}
\paragraph*{SNR Distribution}
We consider a single cell with $N_{UE}$ UEs in connected state.
We assume that the omni-directional DL and UL SNRs for the $i$-th
UE are given by
\beq \label{eq:snr}
    \gamDL_i = \frac{P_{\BSs}}{L_iN_0 W_{\rm tot}}, \quad
    \gamUL_i = \frac{P_{\UEs}}{L_iN_0 W_{\rm tot}}
\eeq
where $P_{\BSs}$ and $P_{\UEs}$
are the transmit powers of the BS and the UE; $L_i$ is the path loss
between the BS and the UE; $N_0$ is the thermal
noise power spectral density (including the noise figure) and $W_{\rm tot}$ is the channel bandwidth.
The path loss values $L_i$ are modeled as independent random variables
with some distribution that depends on the path loss model and
cell radius.  In our evaluation section, we will use the
mmWave statistical path loss model
proposed in \cite{AkdenizCapacity:14}, but the framework can be applied
to arbitrary distributions.
Importantly, the SNRs in
\eqref{eq:snr} are \emph{omni-directional SNRs} in that they do not
include the directional gain.  Depending on the directionality
of the transmissions, we will apply the gains as described in
Section~\ref{sec:SysAssum}. Additionally, losses due to synchronization
errors, beam alignment errors, etc., can be easily incorporated in \eqref{eq:snr}.

\paragraph*{Rounding}  We will assume that all the channels
must be an
integer number of OFDM symbols.  Hence, in many of our calculations, we will need to round up to the smallest number
of integer symbols.  Given an OFDM symbol period $T_{\rm sym}$, let
\beq \label{eq:Qsym}
    Q(T) = T_{\rm sym} \lceil (T/T_{\rm sym}) \rceil,
\eeq
which is the value of time $T$ rounded up to the nearest integer multiple of
$T_{\rm sym}$.

\subsection{Utilization} \label{sec:smallpacket}
Following our discussions in Section \ref{Sec:frameStruct}, we consider analog beamforming based designs for the physical layer data channels. In order to quantify the interplay between the frame structure design choice and the traffic statistics, 
we define the utilization factor $\eta$ as the ratio
between the minimum time required to transmit a particular PDU
and the total time allocated for the corresponding transmission. Specifically,
we analyze the difference between the fixed and the flexible TTI based
designs in terms of $\eta$. We first analyze traffic with TCP ACKs, followed by a more general exposition based on the statistical nature of the data traffic.
%Two important short packet scenarios are investigated:
%traffic with TCP ACKs and short control packets (such as RRC messages).

\paragraph*{Utilization with TCP ACKs}
We consider a simple model with $N$ users, each having one
full buffer TCP flow. A full buffer flow implies that at every
transmission opportunity, the radio resources available for data
transmission are completely used.
Let us assume that we have $N$ such flows in the DL,
though the direction of the flow does not affect the analysis.
 For each of these $N$ flows, the network expects
TCP acknowledgements, which will be transmitted in the UL.
TCP ACKs, unlike the data, have a small size. For example a 1500 byte TCP
data can be acknowledged by a 74 byte ACK. Hence, the ACK will
not need the maximum possible radio resource available and in fact,
the ACKs can be treated as bursty data.

The number of TCP segments $(S_{N})$ that are transmitted in one time slot
for the full buffer transmission is given as
\beq
S_{N} = \frac{T_{\rm TTI, max}~\rho_{\rm DL}W_{\rm tot}}{L_{\rm data}},
\eeq
where $L_{\rm data}$ is the length of each segment in bits,
$\rho_{\rm DL}$ is the spectral efficiency for DL transmission and
$W_{\rm tot}$ is the total bandwidth available for the transmission.

In the reverse direction, each of these $S_{N}$ segments will be acknowledged
by one TCP ACK packet of length $L_{\rm ack}$ which includes all packet
overhead. Thus there are at most $S_N L_{\rm ack}$
bits of data to be transmitted on the UL (we will have less data if ACKs are combined).
The minimum time required to transmit the acknowledgements is
\beq
T_{\rm ack}^{\rm min} = \frac {S_N L_{\rm ack}}{\rho_{\rm UL}W_{\rm tot}},
\eeq
where $\rho_{\rm UL}$ is the UL spectral efficiency. The computation of
the spectral efficiency is presented in the Appendix.

We define $T_{\rm TTI,max}$ as a design specific constant which is the
maximum TTI for a particular mmWave system. For the fixed TTI based
design, all allocations made are for a duration of $T_{\rm TTI,max}$,
regardless of the size of the data to be transmitted.
Thus the utilization factor ($\eta_{\rm fix}$) for the full buffer TCP flow is given by
\beq
\eta_{\rm fix} = \frac {T_{\rm TTI,max}+T_{\rm ack}^{\rm min}}{2T_{\rm TTI,max}}.
\eeq

A flexible TTI based system, on the other hand, will allocate a transmission time
of $T_{\rm TTI,max}$ in the DL but on the UL will allocate an integer
number of symbols based on the packet size.
Thus the total time allocated for the transmission of a TCP ACKs is
$
Q \big( T_{\rm ack}^{\rm min} \big).
$
The utilization factor for flexible TTI based design is,
\beq
\eta_{\rm flex} = \frac {T_{\rm TTI,max}+T_{\rm ack}^{\rm min}}{T_{\rm TTI,max}+Q \big( T_{\rm ack}^{\rm min} \big)}.
\eeq

\paragraph*{Small packets}
As an alternate small packet model, we assume that the number of packets to be transmitted
over the physical channel is a random variable that depends on the
packet generation rate. The size of each of the packets transmitted
during this period is modeled as a random variable ${b}$ with a
known distribution.
The minimum time required to transmit  ${b}$ bits of data in the $i$th time slot is
\beq
T_{i}^{\rm min} = \frac{{b}}{\rho W_{\rm tot}},
\eeq
where $\rho$ is the spectral efficiency of the link
and $W_{\rm tot}$ is the transmission bandwidth.

Following the analysis for the full buffer model, we can write
the time allocated to the bursty flows for both schemes as
\begin{equation}
T_i = \left\{ \begin{aligned}
    T_{\rm TTI,max} & : {\rm Fixed} \\
    Q ( T_{i}^{\rm min} )  & : {\rm Flexible}
  \end{aligned}
\right.
\end{equation}

Unlike the analysis of the first traffic model, it must be noted that the
$T_i$'s are random variables, as they are a function of ${b}$ and $\rho$.
For the fixed TTI based design the expected value of the utilization factor
calculated over a large time duration is given by\footnote{
We consider that packets arriving in a given time interval are served before
packets arrive in the next time interval, i.e., there is no accumulation of packets.}
\beq
\eta_{\rm fix} =\frac{1}{W_{\rm tot}}\Exp\Bigg[\frac{1}{\rho}\Bigg]  \frac{\Exp[{b}]}{T_{\rm TTI,max}}.
\label{eq13}
\eeq
where we note that the length of the MAC PDU is independent of the current spectral efficiency ($\rho$) of the channel. The calculation of $\rho$ is shown in the Appendix.

On the other hand, the steady state utilization factor for the flexible scheme can be found using renewal theory as,
\begin{equation}
\eta_{\rm flex}  =  \lim_{N \to \infty}  \frac{\frac{1}{N}\sum\limits_{i=1}^{N} \left( \frac{b_i}{\rho_i W_{\rm tot}} \right)}{ \frac{1}{N}\sum\limits_{i=1}^{N} Q (\frac{{b_i}}{\rho_i W_{\rm tot}})}
 = \frac{\Exp[{b}] \Exp[1/\rho]}{{W_{\rm tot}}\Exp[Q (\frac{{b}}{\rho W_{\rm tot}})]}.
\label{eq14}
\end{equation}
%\beq
%\eta_{\rm flex} = \frac{1}{W_{\rm tot}}\Exp\Bigg[\frac{1}{\rho}\Bigg]\frac{\Exp[{b}]}{\Exp[Q (\frac{{b}}{\rho W_{\rm tot}})]}.
%\label{eq14}
%\eeq

\subsection{Overhead} \label{sec:overhead}
Based on the discussion on beamfroming architectures in Section \ref{sec:mimo} and control signal multiplexing in Section \ref{sec:digBF}, in this section we analyze the overhead cost of the PHY layer control channel. Detailed analyses for each PHY control message are given as follows.

\paragraph*{Scheduling request}  Assume that there must
be a  dedicated opportunity for each UE to transmit a scheduling
request (SR) at least once every $T_{\rm per,SR}$ seconds.  The value of $T_{\rm per,SR}$
is one component of the UL delay. Let $b_{SR}$ be the number of bits
in the SR and $\gamma_b$ be the minimum $E_b/N_0$ for the channel, where $E_b$ represents the energy per bit.
When the BS has analog BF, there are two options to receive the
SR.  First, it can receive the SRs in a TDMA manner, in which case it gets the
full directional gain on both the BS and the UE side.
Hence we can write
\begin{align*}
& b_{SR}  E_b \leq P_{\rm tx,UL}  T_{per,SR}, \\
{\rm or, \quad} & b_{SR}\left( \frac{E_b}{N_0}\right) \leq \left(\frac{P_{\rm tx,UL}}{N_0 W_{\rm tot}} \right) W_{\rm tot} T_{\rm per,SR},
\end{align*}
where $P_{\rm tx,UL}$ is the UL transmit power and $\left(\frac{P_{\rm tx,UL}}{N_0 W_{\rm tot}}\right)$ is the UL SNR $\gamma^{\rm UL}$.

As the transmit time should be an integer multiple of the symbol length, the time to receive the SR for each UE will be at least
\beq \label{eq:TSRUE}
    T_{\rm SR} = Q\left(
    \frac{b_{\rm SR}\gamma_{\rm b,SR}}{G_{\rm BS}G_{\rm UE}W_{\rm tot}\gamma^{\rm UL}_{\rm min}}\right),
\eeq
where $Q(\cdot)$ is the operator in \eqref{eq:Qsym}. In \eqref{eq:TSRUE}, we have assumed that
the time slot width for the UE is dimensioned for the worst-case
UE and hence we have used the minimum UL SNR, $\gamma^{\rm UL}_{min}$.
When the number of UEs equals $N_{UE}$, the total overhead using TDMA for the SR is:
\begin{align}
     \MoveEqLeft \alpha_{\rm SR,TDMA} = \frac{N_{\rm UE}T_{\rm SR}}{T_{\rm per,SR}} \nonumber
    = \frac{N_{\rm UE}}{T_{\rm per,SR}}Q\left( \frac{b_{\rm SR}\gamma_{\rm b,SR}}{G_{\rm BS}G_{\rm UE} W_{\rm tot} \gamma^{\rm UL}_{\rm min}}\right).
    \label{eq:aSRTDMA}
\end{align}
An alternate option is for the UEs to transmit the SRs simultaneously on different
frequency resources, and have the BS
receive the SRs omni-directionally.
Assuming that the number of degrees of freedom
is sufficient (which is likely since the bandwith is large), the overhead using
this FDMA scheme is
\beq \label{eq:aSRFDMA}
  \alpha_{\rm SR,FDMA} = \frac{1}{T_{\rm per,SR}} Q\left( \frac{b_{\rm SR}\gamma_{\rm b,SR}}{
            G^{\rm omni}_{\rm BS}G_{\rm UE} W_{\rm tot} \gamma^{\rm UL}_{\rm min}}\right),
\eeq
where the BS uses the omni-directional gain.
Finally, for digital BF, the BS can receive all the SR signals simultaneously
while obtaining the directional gain, so the overhead fraction is
\beq \label{eq:bSRFDMA}
  \alpha_{\rm SR,Dig} = \frac{1}{T_{\rm per,SR}} Q\left( \frac{b_{\rm SR}\gamma_{\rm b,SR}}{G_{\rm BS}G_{\rm UE} W_{\rm tot} \gamma^{\rm UL}_{\rm min}}\right).
\eeq

The number of bits $b_{\rm SR}$ will depend on the quantization resolution for the buffer
status.  More bits will be required in the variable TTI mode so that
the BS can schedule the correct TTI size.  Once the UE is scheduled, it can transmit
a full buffer status report inband as in LTE.  Additionally, in either option,
we will need further bits for a CRC to prevent false alarms that would result in UL allocation
wastage.  We will describe this in Section~\ref{sec:eval}.

\paragraph*{UL CQI and other control}
In LTE, in order to enable the BS to perform rate adaptation in the downlink, each UE continuously transmits channel quality indicator (CQI) reports on a dedicated uplink control channel.  The dedicated UL control channel is also used for transmitting indications of the channel spatial rank for MIMO. The overhead for this channel can be computed identically to the SR report above. The resulting expressions will depend on $T_{\rm per,CQI}$, the periodicity of the CQI reports, $b_{CQI}$, the number of bits per report, and $\gamma_{\rm b,CQI}$, the $E_b/N_0$ for the channel.

\begin{figure}
\centering
\subfloat[ ]{\includegraphics [page=2, trim={5cm 14cm 11cm 4.5cm},clip, scale=0.35]{DLGrant.pdf} \label{ULGrantA}}
\hspace{0.25cm}
\centering
\subfloat[ ]{\includegraphics [page=3, trim={5cm 14cm 9cm 4.5cm},clip, scale=0.35]{DLGrant.pdf} \label{ULGrantD} }
\centering
\subfloat[ ]{\includegraphics [page=1, trim={5cm 14cm 9cm 4.5cm},clip, scale=0.35]{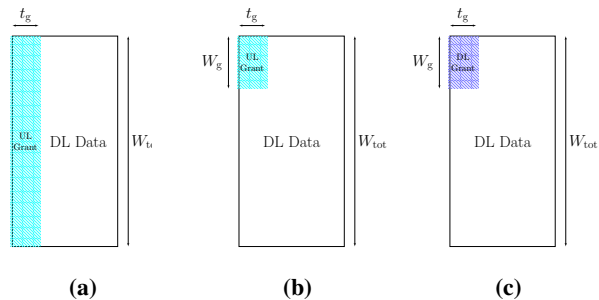} \label{DLGrant} }
\caption{(a) UL grant placing for analog beamforming. (b) UL grant placing for digital/hybrid beamforming. (c) DL grant multiplexed with DL data for both digital and analog architectures.}
\label{Grants}
\end{figure}

\paragraph*{UL grants}
The overhead taken by the UL grants depends critically
on the MIMO architecture.  If the BS has only analog BF, it
can transmit in only one direction at a time, and therefore cannot
transmit a UL grant to one UE and data to another UE simultaneously.
Thus, the UL grant must be transmitted by itself across the entire
bandwidth as shown in Fig. \ref{ULGrantA}.  If we let $b_{\rm g}$
be the number of bits in the grant, and $\gamma_{\rm b,g}$ be
the minimum $E_b/N_0$, the minimum time per grant will be
\begin{equation}
    T_{\ULG,\rm ana} =
    Q\left( \frac{b_{\rm g}\gamma_{\rm b,g}}{\GBS\GUE W_{\rm tot} \gamDL_i}\right).
\end{equation}
Now, say that a fraction $p_\ULs$ of the TTIs is allocated for the
UL and $1-p_\ULs$ for the DL.  Since there is one
UL grant for every TTI allocated to the UL, the overhead for
the UL grants with analog BF will be
\begin{equation} \label{eq:aULgAn}
    \alpha_{\ULG,\rm ana} = \frac{p_{\ULs}}{\Exp\left[ \TTTI \right]} \Exp \left[ Q\left( \frac{b_{\rm g}\gamma_{\rm b,g}}{\GBS\GUE W_{\rm tot} \gamDL_i}\right)\right],
\end{equation}
where the expectation in $\Exp [ \TTTI]$ is taken over the TTI sizes,
and the expectation in $\Exp[ Q(\cdot)]$ is taken over the variability
in the DL SNR.  Note that in the fixed TTI case $\TTTI$ is a
fixed value.

When the BS has hybrid BF with $K>1$ streams or digital BF,
the BS can transmit the UL grant in a fraction of the bandwidth
while transmitting DL data
to other UEs in the remainder of the bandwidth
as shown in Fig. \ref{ULGrantD}.
To evaluate the overhead in this case, suppose that the UL grant
is sent for $T_{\rm g}$ seconds over a bandwidth of $W_{\rm g}$.
If the DL power is allocated uniformly across the total bandwdith
$W_{\rm tot}$, then the grant will be received at UE $i$ with an SNR of
%$\frac{P_{i}^{DL}W_{\rm g}/W_{\rm tot}}{LN_0W_{\rm tot}} =
$\gamDL_i$.
Hence, the minimum transmission time for the grant
with digital or multi-stream hybrid beamforming will be
\[
    T_{\ULG,\rm dig} =
    Q\left( \frac{b_{\rm g}\gamma_{\rm b,g}}{\GBS\GUE  \gamDL_i W_{\rm g}}\right),
\]
and the overhead will be
\beq \label{eq:aULGdig1}
    \alpha_{\ULG,\rm dig} = \frac{W_{\rm g} p_{\ULs}}{W_{\rm tot}\Exp\left[ \TTTI \right]} \Exp \left[Q\left( \frac{b_{\rm g}\gamma_{\rm b,g}}{\GBS\GUE W_{\rm g}\gamDL_i}\right)\right].
\eeq
Now, in principle, $W_g$ can be adjusted so that there is no rounding error
in the $Q(\cdot)$ function, i.e., $Q(x) = x$.  In this case,
\eqref{eq:aULGdig1} simplifies to
\begin{equation} \label{eq:aULGdig}
    \alpha_{\ULG,\rm dig} = \frac{p_{\ULs}b_{\rm g}\gamma_{\rm b,g}}{\GBS\GUE W_{\rm tot} \Exp\left[ \TTTI \right]} \Exp\left[  \frac{1}{\gamDL_i}\right],
\end{equation}
which is the same expression as in the analog case \eqref{eq:aULgAn}, but
without the quantization. In fact, we remark that the additional overhead incurred
by the analog beamforming architecture is due to the difference between $T_{\rm ULG,ana}$
and its quantized value.

\paragraph*{DL grants}
Since the DL grant and corresponding DL data are transmitted to the same UE,
they can be multiplexed together
as shown in Fig.\ \ref{DLGrant}. This multiplexed transmission can be
 performed for all the beamforming architectures.
Thus
following the analysis provided for the UL grant, we can express the overhead due to the DL grant as
\begin{equation} \label{eq:aDLG}
    \alpha_{\DLG} = \frac{(1-p_{\ULs})b_{g}\gamma_{\rm b,g}}{\GBS\GUE W_{\rm tot} \Exp\left[ \TTTI \right]} \Exp\left[  \frac{1}{\gamDL_i}\right].
\end{equation}

\paragraph*{DL and UL ACKs}
The DL ACK is sent by the BS in response to UL data received from the UE.
Its time-frequency allocation has the same
constraints as the UL grant.  Applying a similar derivation we obtain that
with analog beamforming the DL ACK overhead is given by
\beq \label{eq:aDLAAn}
    \alpha_{\DLACK,\rm ana} = \frac{p_{\ULs}}{\Exp\left[ \TTTI \right]} \Exp \left[Q\left( \frac{b_{\ACK}\gamma_{\rm b,\ACK}}{\GBS\GUE W_{\rm tot} \gamDL_i}\right)\right],
\eeq
where $b_\ACK$ is the number of bits per ACK and $\gamma_{\rm b,\ACK}$ is its $E_b/N_0$
requirement.  Note that the number of bits may be greater than one due to separate ACKs
in different spatial streams as in 3GPP or in the case when subunits within the MAC PDU
are ACK-ed individually.
Similarly for the case of hybrid beamforming ($K$ streams) or
fully digital beamforming, the overhead is
\begin{equation} \label{eq:aDLADig}
    \alpha_{\DLACK,\rm dig} = \frac{p_{\ULs}b_{\ACK}\gamma_{\rm b,\ACK}}{\GBS\GUE W_{\rm tot} \Exp\left[ \TTTI \right]} \Exp\left[  \frac{1}{\gamDL_i}\right].
\end{equation}

The DL ACK is sent by the UE in response
to DL data and has similar time-frequency options as the UL ACK,
but different overhead.
In the analog BF case, it must be sent by itself and the overhead is
\beq \label{eq:aULAAn}
    \alpha_{\ULACK,\rm ana} = \frac{1-p_{\ULs}}{\Exp\left[ \TTTI \right]} \Exp \left[Q\left( \frac{b_{\ACK}\gamma_{\rm b,\ACK}}{\GBS\GUE W_{\rm tot} \gamUL_i}\right)\right].
\eeq
In hybrid BF with multiple digital streams or in digital BF,
the UL ACK can be multiplexed with UL data from other UEs.
In this case, the UE can transmit all its power on the ACK, and
the ACK will only be bandwidth limited.
Thus, the allocation will only be limited by the spectral efficiency
of the UL ACK.  Suppose that the UL ACK is transmitted at a spectral efficiency
$\rho_{\ACK}$.  Then, the UL ACK will require $b_\ACK/\rho_\ACK$ degrees of freedom
to transmit.  In a period of $\TTTI$ seconds and bandwidth $W_{\rm tot}$, there
are a total of $\TTTI W_{\rm tot}$ degrees of freedom.  So, the
UL ACK overhead with multi-stream hybrid BF or digital BF is given by
\beq \label{eq:aULAAn}
    \alpha_{\ULACK,\rm dig} =
    \frac{(1-p_{\ULs}) b_\ACK}{\rho_\ACK \Exp\left[ \TTTI \right]W_{\rm tot}},
\eeq
which will be negligible in the bandwidths of interest for mmWave systems.

\section{Evaluation for Realistic Design Scenarios} \label{sec:eval}

Following our analysis for utilization and control overhead, in this section
we evaluate the beam forming and frame design options. We begin with a discussion on
the selection of realistic system parameters in Subsection \ref{sec:SysAssum}. This is followed by a detailed analysis of the frame structure and beamforming choices based on the utilization and overhead parameters in Subsection \ref{evalres}. Finally we provide a concise summary of our findings in Subsection \ref{summary}

\subsection{System Assumptions} \label{sec:SysAssum}
\paragraph{System Parameters}
\begin{table}
\centering
\begin{tabular} {|c|p{9cm}|p{0.7cm} |p{0.75cm}|p{0.6cm}|}
\hline
\multirow{2}{*}{Parameter} & \multirow{2}{*}{Description}& \multicolumn{2}{c|}{Value used}& \multirow{2}{*}{Units} \\ \cline{3-4}
& & Fixed & Flexible & \\ \hline
$\gamma_{\rm b,M}$ & SNR per bit ($E_b/N_0$)  for a given control message M &  \multicolumn{2}{c|}{6} & dB \\ \hline
%$\gamma_{i}^{\rm DL}$ & SNR for the $i^{th}$ downlink channel without beamforming & -12.0 & dB \\ \hline
%$\gamma_{i}^{\rm UL}$ & SNR for the $i^{th}$ uplink channel without beamforming& -18.1 & dB \\ \hline
$G_{\rm BS}$ & Directional beamforming gain for the base station &  \multicolumn{2}{c|}{18} & dB \\ \hline
$G_{\rm BS}^{\rm Omni}$ & Max beamforming gain for the BS assuming
 $N^{\rm ant}_{\rm BS}=64$.  &  \multicolumn{2}{c|}{0} & dB \\ \hline
$G_{\rm UE}$ & Max beamforming gain for the user equipment.
 Assume $N^{\rm ant}_{\rm UE}=16$.  &  \multicolumn{2}{c|}{12}& dB \\ \hline
$T_{\rm TTI}$ &Transmission time interval & \multicolumn{2}{c|}{125} & $\mu$s \\ \hline
$T_{\rm sym}$& Duration of one OFDM symbol &  \multicolumn{2}{c|}{ 4.16} & $\mu$s \\ \hline
$T_{\rm per,SR}$& Period in which all UEs can transmit a SR at least once &  \multicolumn{2}{c|}{ 500} & $\mu$s \\ \hline
$W_{\rm tot}$ &Total system bandwidth &  \multicolumn{2}{c|}{1000} & MHz \\ \hline
%$W_{g}$ & Bandwidth allocated for grants & 100 & MHz \\ \hline
$\rho_{\rm ACK}$ & Spectral efficiency for ACK transmission &  \multicolumn{2}{c|}{$\frac{1}{8}$} & bps/Hz  \\ \hline
$p_{\rm UL}$ & Fraction of uplink packets &  \multicolumn{2}{c|}{0.5} & \\ \hline
$b_{\rm SR}$ &Size of a scheduling request& 18 & 26 \quad 42  & bits \\ \hline
$b_{\rm g} $&Size of a UL/DL grant& 80& 100 & bits \\ \hline
$b_{\rm ACK} $&Size of a HARQ acknowledgement&  \multicolumn{2}{c|}{5} & bits \\ \hline
\end{tabular}
\caption{Parameters used for the system evaluation.}
\label{Tab:Param}
\end{table}

We leverage the above utilization and control overhead analysis
to evaluate the different frame structure and signaling options
under realistic design scenarios.
The parameters are detailed in Table~\ref{Tab:Param}.
Following the capacity analysis in \cite{AkdenizCapacity:14},
we assume $N^{\rm ant}_{\rm BS}=64$ antennas at the BS
and $N^{\rm ant}_{\rm UE}=16$ at the UE -- reasonable dimensional
arrays for mmWave systems.
The number of users connected to the BS is given by $N_{\rm UE}$ and is varied. 
%We vary $N_{\rm UE}$ for one BS to capture the effect it has on the control overhead of the system using the parameters
%in Table \ref{Tab:Param}.

 \paragraph{SNR Distribution}

\begin{figure}[!t]
\centering
\includegraphics [scale=0.4]{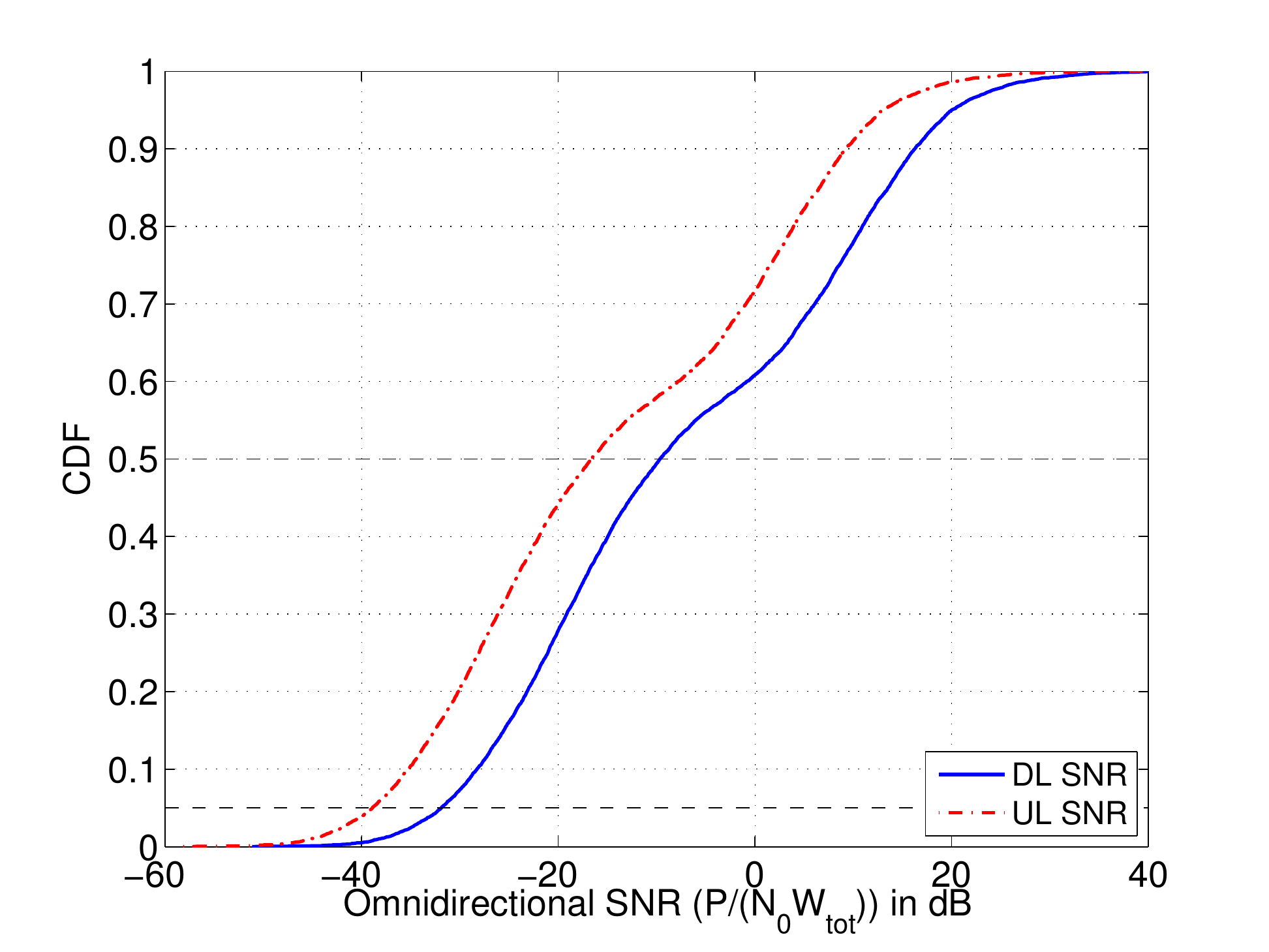}
\caption{Distribution of DL and UL SNR without beamforming gains for users distributed uniformly in a cell of radius 100 m using the path loss model in \cite{AkdenizCapacity:14}.}
\label{figsnr}
\end{figure}
The transmit powers in the DL and UL directions are taken to be 30 dBm and 20 dBm respectively. The noise
figures are 7 dB and 4 dB in DL and UL, respectively, consistent
with the capacity analysis in \cite{AkdenizCapacity:14}.
We used a path loss model in \cite{AkdenizCapacity:14} derived from
actual measurements at 28~GHz in New York City~\cite{rappaportmillimeter}.
The distributions of the SNRs $\gamma_{i}^{\rm DL}$ and
$\gamma_{i}^{\rm UL}$ are then generated from this model under the assumption
that the mobiles are uniformly located in the cell radius of 100~m from the
base station.  The resulting distribution is shown in
Fig.~\ref{figsnr}
along with the 5\% and median lines.
For the UL the 5\% and the median SNRs are
$-39$ dB and $-16$ dB respectively. For the DL, the respective values are $-32$ dB and $-9$ dB.  The 5\% values are used as the target SNRs for the minimum SNRs
$\gamma_{\rm min}^{DL}$ and $\gamma_{\rm min}^{UL}$.

\paragraph{OFDM Symbol Period}
The OFDM symbol period is critical in determining the overhead for transmitting small control and data messages -- a key aspect of the frame structure design.
Very short OFDM symbols allow time to be divided into small intervals enabling small data packets to be transmitted with minimal padding.  On the other hand, each OFDM symbol contains a cyclic prefix (CP) whose size
is determined by the channel delay spread and synchronization errors.
Reducing the OFDM symbol period increases the percentage overhead incurred due to the CP.
In this work, we use the OFDM parameters proposed in \cite{khan2012millimeter}.  
The authors propose an OFDM symbol duration is 3.70 $\mu$s with a CP duration of 0.463 $\mu$s for small cells
($< 1$ Km in radius), giving a total symbol period of 4.16 $\mu$s. This CP duration is sufficient for delay spreads
as measured in \cite{Samimi2015Prob,MacCartney2015Wideband}.
The work in \cite{khan2012millimeter} also uses a fixed TTI with 30 OFDM symbols
corresponding to $T_{\rm TTI}$ of 125 $\mu$s.  We use this value as the fixed TTI. 

%Other works, \cite{levanen2014dense} and  \cite{levanen2014radioaccess}, propose alternative design choices
%which can be easily substituted in our model.

The values of the other parameters in Table~\ref{Tab:Param} are justified in the Appendix.

\subsection{Evaluation Results}\label{evalres}

\subsubsection{Control Overhead}

Following the analysis in Section \ref{sec:overhead}, we compare the overheads due
to the physical layer control signals for analog, hybrid and fully digital beamforming
architectures.
As an example, the overhead due to
the various control signals when a BS serves 8 users is listed in Table \ref{tabOH}.
We note that for analog beamforming the overhead is around 12\% while
with a $K=2$ hybrid architecture the overhead dips to 3\%.
We notice for both these cases that the SR dominates the overhead.
For the low power fully digital architecture it is less than 1\%. Thus, as discussed in
Section \ref{sec:digBF}, the overhead is considerably reduced when hybrid or fully
digital beamforming is used.

\begin{table}
\centering
\begin{tabular} {|p{4cm}|p{1cm}|p{1cm}|p{1cm}|p{2cm}|p{1cm}|}
\hline
\multirow{3}{3cm}{Control Message}& \multirow{3}{1.5cm}{Message Type} & \multicolumn{4}{c|}{Overhead} \\ \cline{3-6}
& &\multicolumn{2}{c|}{Analog} & Hybrid ($K$=2) & Digital \\ \cline{3-4}
& &TDMA &FDMA & &\\ \hline
\multirow{3}{3cm}{Scheduling Request} & Trigger& 0.0667 & 0.0750 &0.0333 &0.0083 \\  \cline{2-6}
& Short & 0.0667 &0.1083  & 0.0333& 0.0083 \\  \cline{2-6}
& Long & 0.0667 & 0.1667 &0.0333 &0.0083 \\ \hline
Uplink Grant & & 0.0167 &  N.A & 0.000177 & 0.000184\\ \hline
Downlink Grant & &0.000177 & N.A & 0.000177&0.000177\\ \hline
\multirow{2}{3cm}{HARQ ACK} & DL & 0.0167 & N.A & 0.000009 &0.000009 \\ \cline{2-6}
& UL  & 0.0167& N.A & 0.00016 &0.00016 \\\hline
\multicolumn{2}{|c|}{\bf Total} & \multicolumn{2}{c|}{\bf 0.1170} & {\bf 0.0339} & {\bf 0.0089} \\ \hline
\end{tabular}
\caption{Control message overheads for the various design alternatives with $N_{UE}$= 8.}
\label{tabOH}
\end{table}

\begin{figure}[!t]
\centering
\includegraphics [scale=0.42]{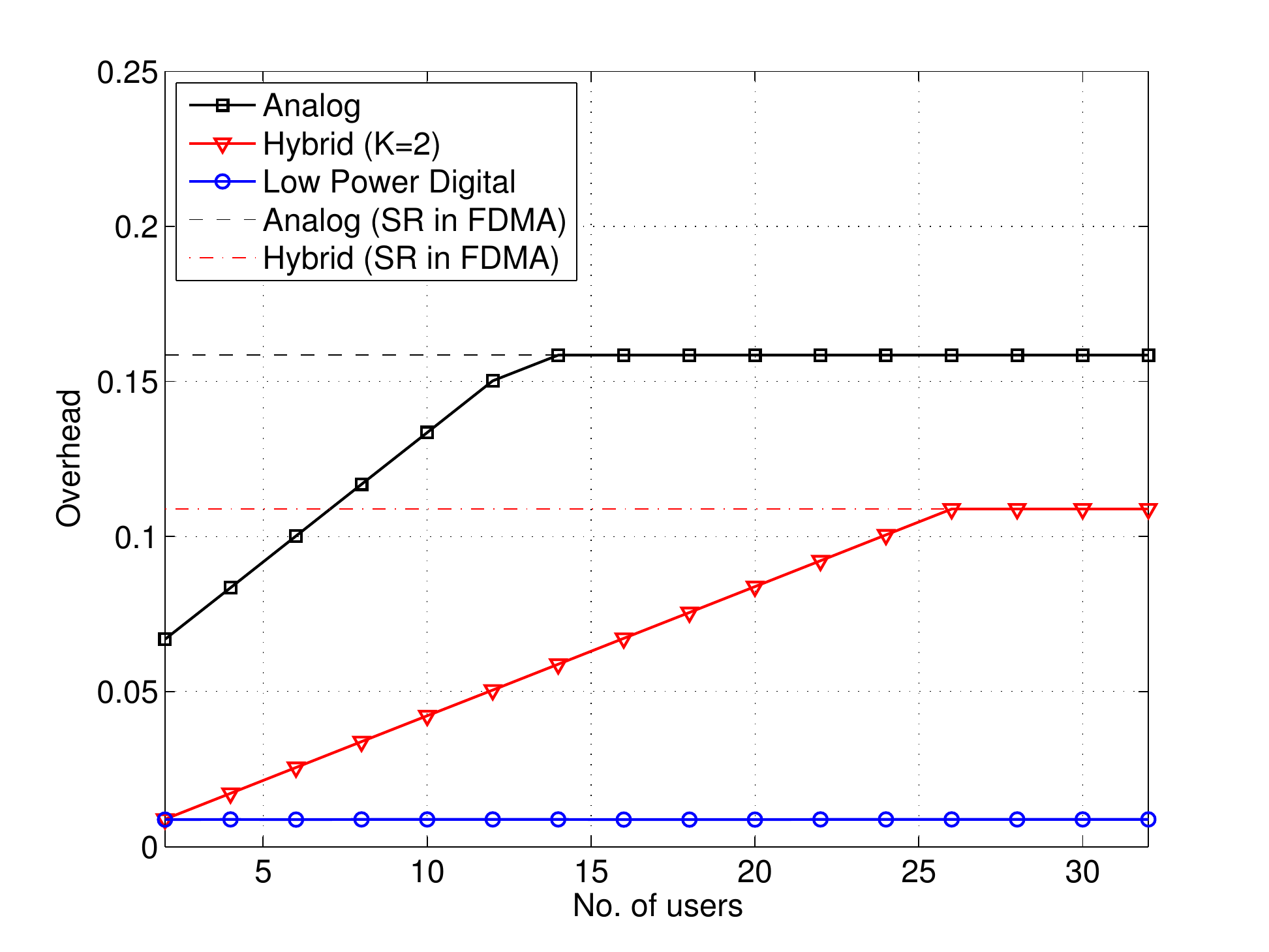}
\caption{Control Overhead versus the number of users for analog, hybrid $(K=2)$ and digital BF architectures.}
\label{Fig:OH}
\end{figure}

\begin{figure*}
\subfloat[ ]{\includegraphics [height=0.27\textwidth, width=0.32\textwidth]{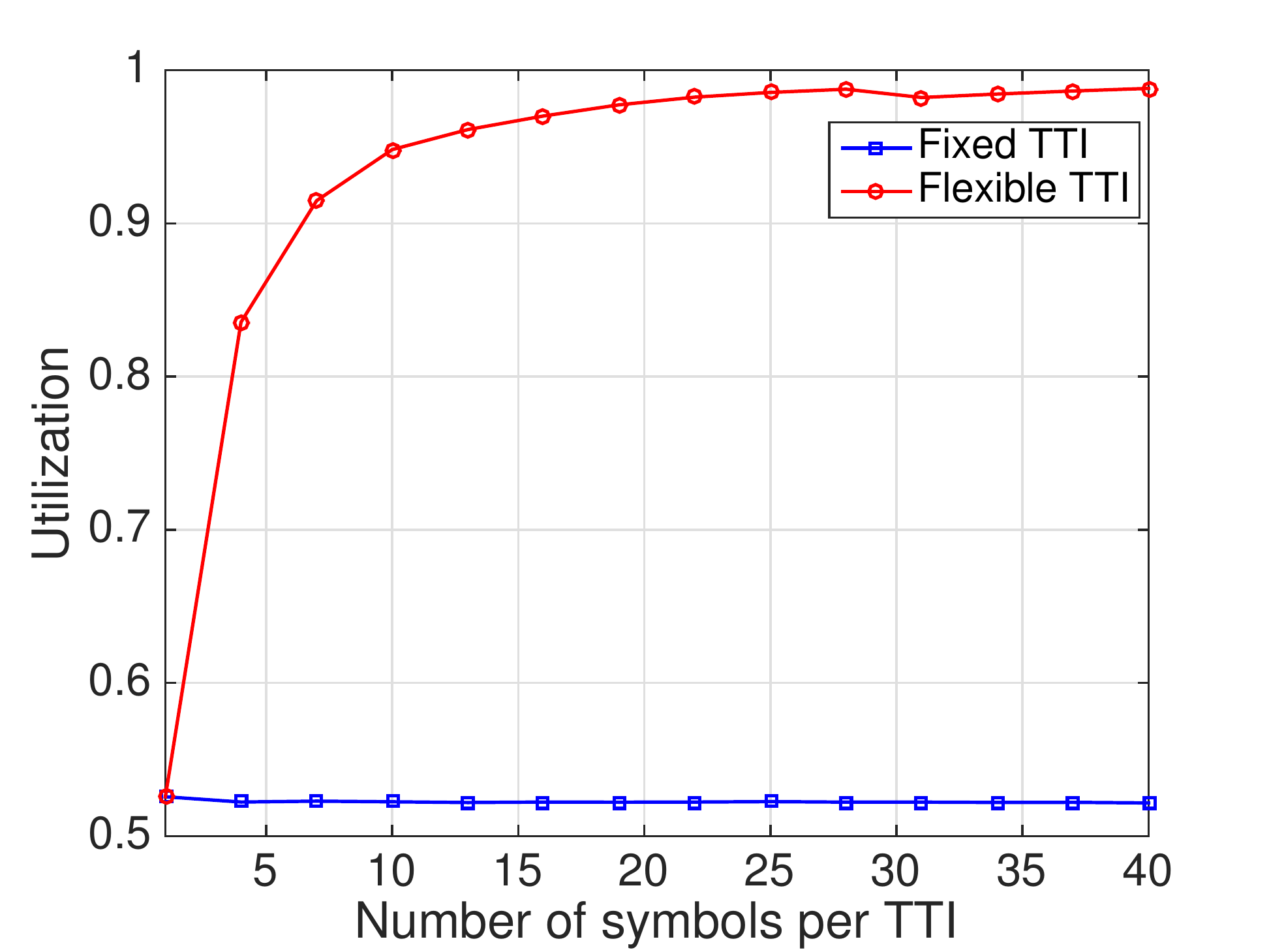} \label{fig:utilizationTCP}}
\subfloat[ ]{\includegraphics [height=0.27\textwidth,width=0.32\textwidth]{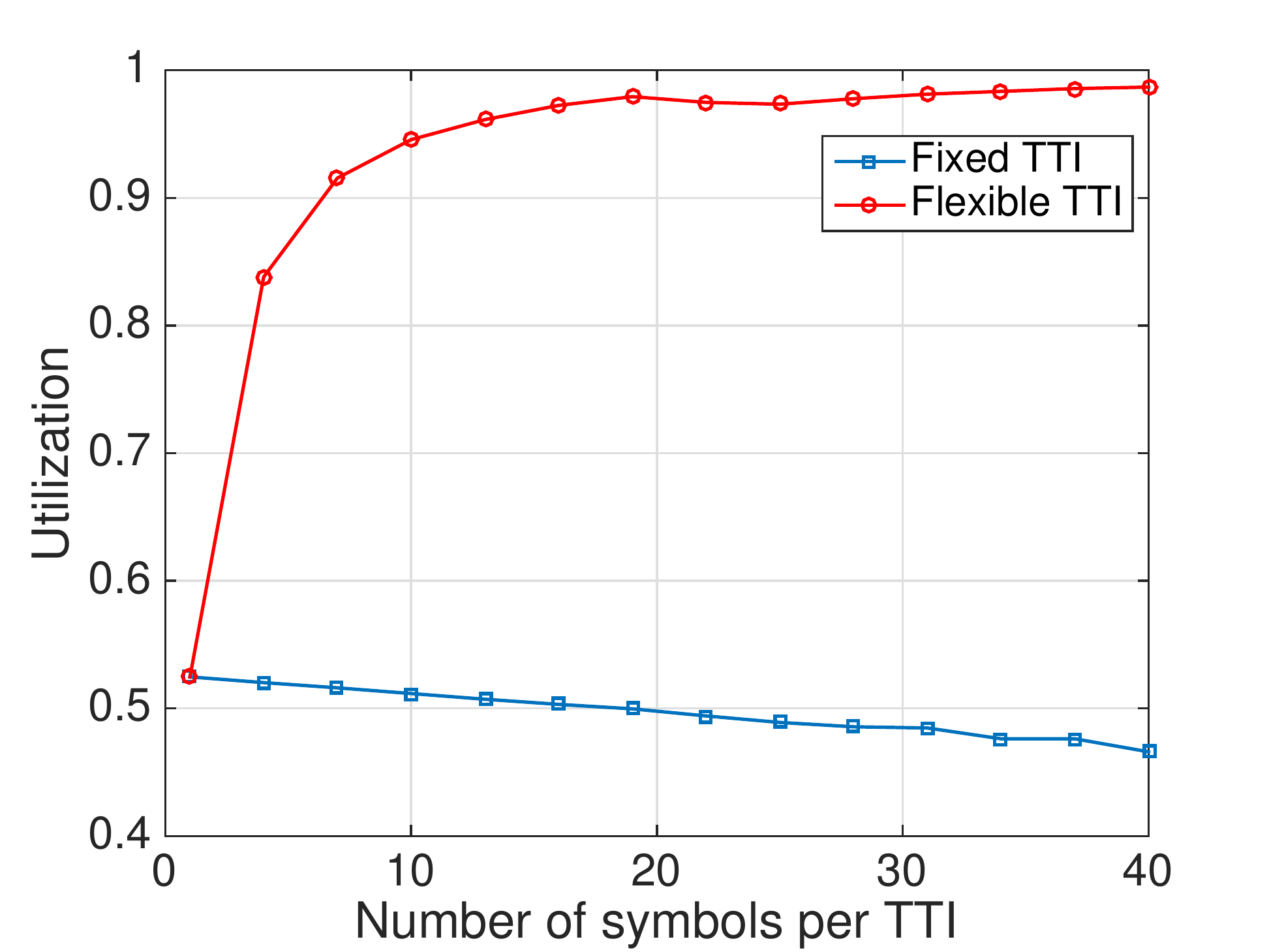} \label{fig:utilizationLargePac} }
\subfloat[ ]{\includegraphics [height=0.27\textwidth, width=0.32\textwidth]{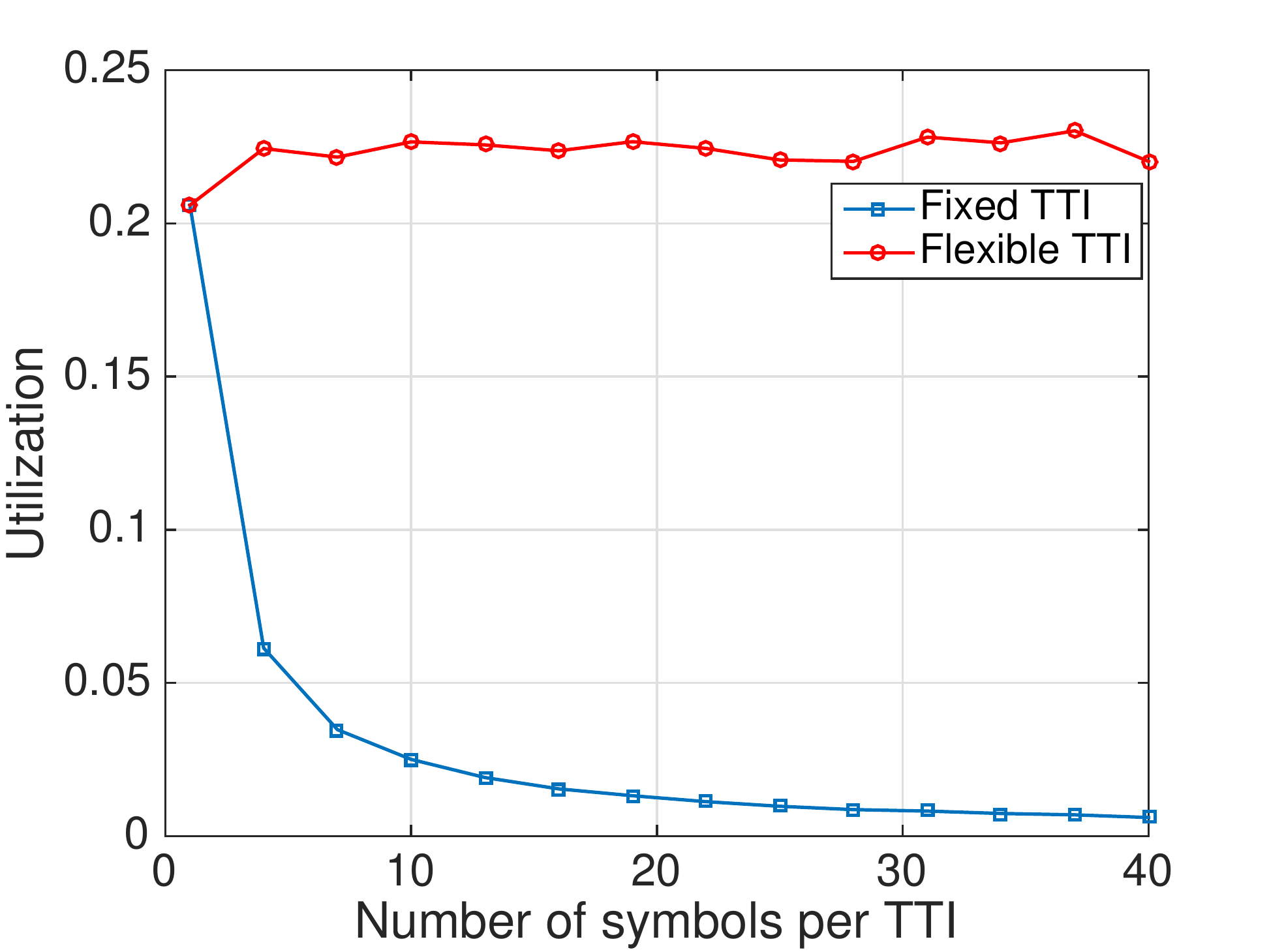} \label{fig:utilizationSmallPac} }
\caption{Utilization vs. maximum TTI with symbol length 4.16 $\mu$s for fixed and flexible
TTI based frame design with (a) full buffer data, (b) large packets (0.5 MB to 5 MB) arriving at a rate 1 per second
 (c) packets between 100 B and 2 MB arriving at a rate of 5 per second. ($N_{UE} = 32$).}
\label{UtilizatioFigs}
\end{figure*}

Fig. \ref{Fig:OH} plots  the overhead as a function of the number of users served by
the BS.
The linear increase of the overhead with the number of users is
attributed to the increase in UL SRs.
In Fig. \ref{Fig:OH} the switch to FDMA based transmission for SRs
 accounts for the plateau reached by the curves for analog and hybrid beamforming.
 Thus we can say that when the number of users is high, analog or hybrid beamforming
 based systems should employ FDMA for UL SR transmission.
We note that for analog beamforming based architectures, even for a smaller
number of connected users, the overhead is considerably higher than for digital
or hybrid architectures. Moreover, for fully digital architecture, the overhead
is constant even when the number of connected users grows.
It should be noted here that this gain in overhead comes at the price of increased
hardware complexity and power consumption for the hybrid and the fully digital
architectures. In order to limit the power consumption we use low resolution ADCs
for the fully digital scheme, which account for the adjustment in effective SNR as
given by \eqref{eq:snrquant}.

\subsubsection{Utilization} Next we compare the utilization of allocated radio resources for the fixed and variable TTI
designs based on the analysis in Section~\ref{sec:smallpacket}.

\paragraph*{Utilization with TCP ACKs} 
Fig.~\ref{fig:utilizationTCP} captures the effect of the maximum TTI ($T_{\rm TTI, max}$)
on the utilization for the full buffer TCP model. In this figure the x-axis shows the number of symbols in $T_{\rm TTI, max}$. We note that, for a given symbol duration, the number of symbols times the symbol duration equals $T_{\rm TTI, max}$.
We see that the fixed TTI scheme gives a constant utilization of around
53\% regardless of the TTI size, implying a dramatic
wastage in bandwidth.  This loss occurs since although the TCP ACK packets
are much smaller that the TCP data packets,  in the fixed TTI mode, both
are transmitted over the same TTI. Hence most the resources allocated
for the TCP ACKs are essentially wasted. In contrast, with the flexible TTI scheme, increasing $T_{\rm TTI, max}$
enables more TCP data packets to be transmitted
per TTI on the forward link, which results in more ACKs in the reverse link. But as ACKs are
much smaller than the data, these can be transmitted over a few symbols.
Thus for the flexible TTI scheme, the utilization is comparable with the fixed design
when $T_{\rm TTI, max}$ is small but rapidly improves as $T_{\rm TTI, max}$ increases.

The relevance of the result lies in the fact that one might argue that using a small value of $T_{\rm TTI, max}$ for the fixed TTI design will ensure very little loss in the utilization. On the contrary, we see that even when a frame is designed with as low as 4  symbols per slot, the resource utilization offered by the flexible scheme is considerably greater. Although one may claim that such a result is qualitatively to be expected, an exact quantification of the gain achieved by using the flexible TTI based frame has never been reported in the literature so far.

As a second test for utilization, based on \eqref{eq13} and \eqref{eq14}, we compute the resource utilization in the case when the MAC PDUs have large size but arrive at a slow rate. The number of PDUs available for a user in each time interval is modeled as a Poisson random variable with mean 1 packet/s. The sizes of the PDUs (in bytes) are truncated log-normal random variables, between 0.5 MB to 5 MB, with mean PDU size of 2 MB and standard deviation 0.722 MB. Each delivered packet is acknowledged
with an ACK by the receiver. Fig. \ref{fig:utilizationLargePac} plots the utilization versus $T_{\rm TTI, max}$ for $N_{\rm UE}= 32$.
For the flexible TTI based scheme the trend is similar to the TCP full buffer case. Conversely, with increasing $T_{\rm TTI, max}$ the utilization for the fixed TTI scheme degrades, as for larger values of $T_{\rm TTI, max}$, even large data packets drawn from this distribution will not be able to fully utilize the allocated resources.

We next analyzed the utilization with small packet sizes and high average arrival rate.
We used the model in \cite{cdma2000}
 with an arrival rate of 5 packets/s, where
the packet sizes are truncated log-normal random variables with mean PDU size  10710 B and standard deviation 25032 B.
The  packet
sizes generated are between 100 B and 2 MB. Fig. \ref{fig:utilizationSmallPac} shows the variation of the utilization with
$T_{\rm TTI, max}$ for this scenario. We observe that for fixed TTI based design the utilization decreases rapidly from
20\% to 1\% as $T_{\rm TTI, max}$ increases. For flexible TTI based design, the utilization remains somewhat constant
around 22\%. This is because in this case, most of the time only one symbol is allocated for data transmission in both directions,
and thus a constant amount of allocated radio resource is being utilized on average.

Thus we see that a variable TTI-based design offers significantly improved
 utilization in comparison to a fixed TTI design in networks
where data packets are short and
bursty. Aggregating packets over multiple arrivals for the same UE
may mitigate this problem and allow us to
use the fixed design. However, 
%we must note that this approach will severely affect the latency of the system and
%will not be acceptable for packets with strict deadlines. Moreover, 
aggregating small packets to saturate an entire subframe
might require a wait time longer than what is acceptable by most applications.

\begin{figure}[t!]
\centering
\includegraphics [scale = 0.42]{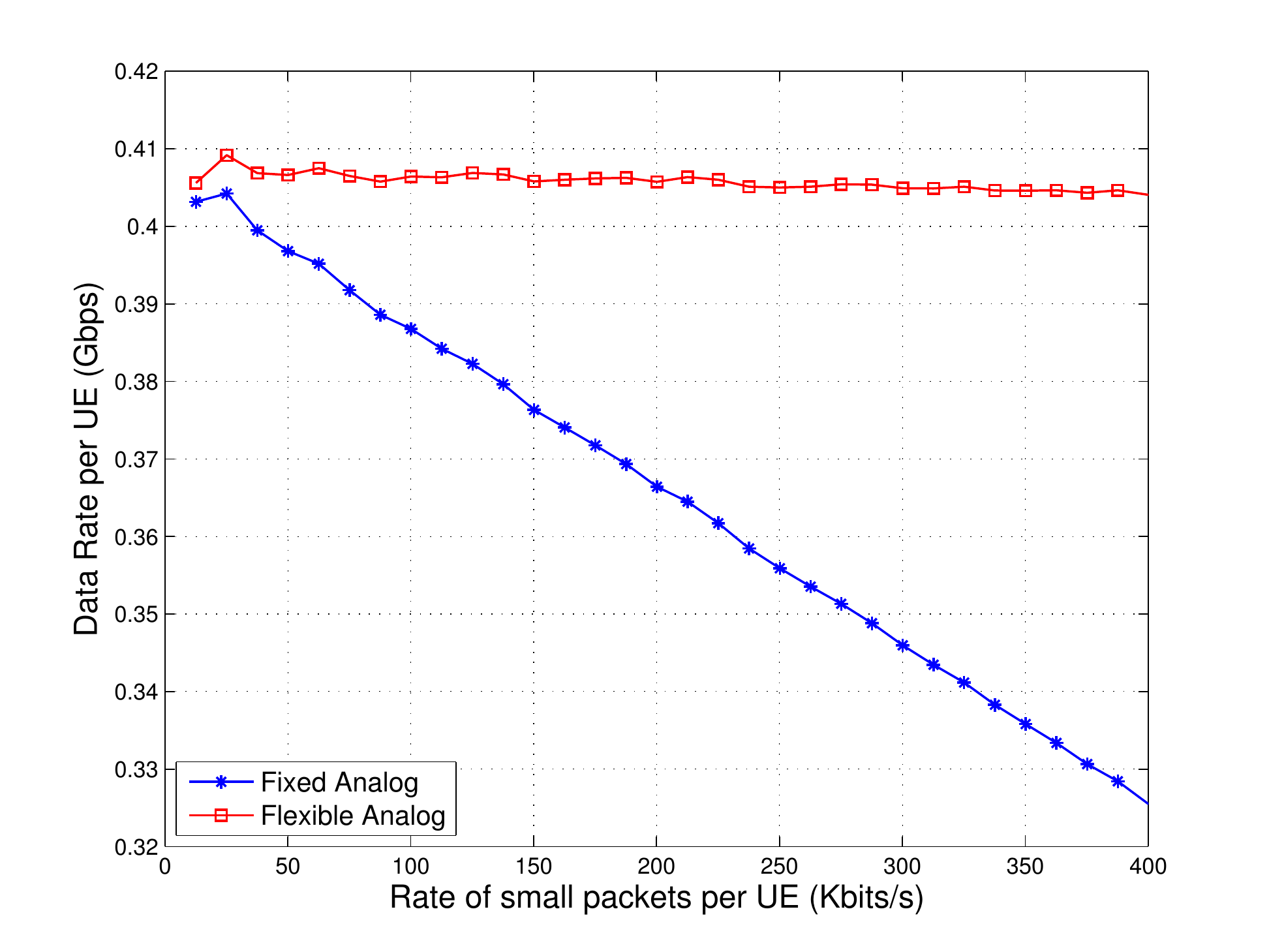}
\caption{The effect of RRC packets on the data rate of the
users for fixed and flexible TTI based designs ($N_{UE} = 8$).}
\label{fig:utilization}
\end{figure}

\paragraph*{Utilization with small control messages}
Another important source of small MAC PDUs arises for RRC control messages.  RRC messages will be used for a variety of
control signaling including interference coordination, measurement reports,
resource allocation, etc.  5G mmWave systems will likely need a greater
rate of control signaling to handle the more rapid fluctuations in the channel.
To analyze this situation,
we consider the case when the BS is serving
$N_{\rm UE}=8$ users.
Each UE is transmitting full buffer traffic in the UL or the DL.
At the same time, the BS sends small control messages to each UE
at a fixed rate.
Similar to current RRC messages
\cite{3GPP36.331}, we assume that each control message is 2000 bits long.
The variable in this experiment is the rate at which the RRC messages are transmitted.
To show the effect of the RRC control messages,
in Fig. \ref{fig:utilization} we plot the data rate achieved be each user versus the rate of RRC control messages.
It can be seen that for the fixed TTI based scheme
the data rate falls rapidly with the increase in the rate of RRC messages.
However, for the flexible
TTI scheme, the data rate decreases very slowly (if at all).

This experiment also gives some insight on the effect that short machine type communications (MTC)
\cite{shariatmadari2015machine} will have on networks. MTCs are characterized by short data packets
(in the order of 100 bytes) with high priority. A key goal for 5G systems is to incorporate a large
number of  MTC devices. From our experiment we can infer that if the fixed TTI based
design is used, the data rates of human-to-human links will degrade considerably in the presence of MTC. On the other hand, systems using a flexible TTI frame can transmit machine type packets with a negligible effect on user experience.

\subsection{Summary of the key findings} \label{summary}
Our analysis demonstrates several appealing features of the proposed design:
\begin{itemize}
\item \emph{Support of low-latency communication:}  In the proposed design, 
the dominant delay is the periodicity of the SR.
Specifically, before transmitting in the UL, the UE may have to wait for a SR opportunity.
Our design has set this to 500~$\mu$s.  Since all other transmissions can occur within a single OFDM symbol of 4~$\mu$s,
it is possible for this design to enable sub-one millisecond round-trip time.  However, the specific time
will need further analysis as it depends on the processing and scheduling times at the mobile and base station.

\item \emph{Support of small packets:}  We have shown that the utilization of the link with a fixed TTI 
is very poor in the presence of small packets.  Such small packets can occur in TCP ACKs as well as control signaling.
In contrast, the proposed flexible design is able to support short packets with minimal overhead loss.

\item \emph{Further benefits with digital beamforming:}  Most current mmWave designs have 
assumed analog beamforming where the BS can ``look" in only one direction at time.
This limits the multiplexing capabilities significantly.  We show that a low-resolution fully digital design
can multiplex users, obtaining significantly better overhead, while using comparable power.
\end{itemize}

\section{Conclusions and Future Work}\label{sec:conc}

Suitable frame structures for mmWave cellular systems will need to support very low latencies and large numbers of users per cell while relying on highly directional transmissions.  We have proposed a novel frame structure
that departs in several key ways from current 4G LTE as well as recently proposed 5G mmWave system design.
Most importantly, in the proposed system, the data and control channels can be scheduled
dynamically in highly granular locations thereby enabling very low latency and the ability to accommodate
mixtures of short and large MAC PDUs.  Different multiplexing schemes are described 
depending on the front-end constraints of the MIMO transceiver, which determines the level of possible
multiplexing.  We have presented an analytic framework to evaluate the system under various
statistical models for the traffic, SNRs, and control periodicity.  This model was then 
applied to realistic system parameters to assess the feasibility of the design in practical scenarios.

Nevertheless, further analysis is warranted.  Our design has abstracted out much of the control signaling,
link-layer aspects and processing capabilities.  These will need to be designed and evaluated.
The latency analysis in particular will depend on the hardware processing capabilities and the detailed
traffic model.  Also, our model has assumed connectivity between a UE and BS.  It is widely-believed that
mmWave systems will use relays and also schedule traffic from multiple cells.  The current MAC-layer design
has not incorporated either of these aspects, which remain as interesting areas for future investigation.

\appendix[Simulation Parameter Selection Details] \label{Appendix}
In this section we delineate the logic behind the selection of the system parameters used
for the simulations and the results. Moreover, we also illustrate some of the considerations that
should be made in selecting these values for a practical system.

\paragraph{Antenna pattern} We assume a set of two dimensional antenna arrays at both the BS and the UE.
On the BS side, the array is comprised of $8 \times 8$ elements and on the UE side we have $4 \times 4$ elements.
The spacing of the elements is set at $\lambda /2$, where $\lambda$ is the wavelength.
These antenna patterns were considered in \cite{AkdenizCapacity:14} and shown to offer excellent
system capacity for small cell urban deployments. In addition, a $4 \times 4$ array operating in the
$28 \; \text{GHz}$ band, for instance, will have a size of roughly $1.5\; \text{cm} \times 1.5 \; \text{cm}$.
The maximum gain that can be achieved by beamforming with an $N_t$ element antenna array, as pointed out in
\cite{AkdenizCapacity:14}, is given in dB as $10\log_{10} N_t$. Thus for the $8 \times 8$ elements array
at the base station the maximum beamforming gain is $10\log_{10} 64 = 18$ dB. For the UE the maximum beamforming
gain is $12$ dB.

\paragraph{Spectral Efficiency}
The spectral efficiency, $\rho$, for a given channel is given in \cite{AkdenizCapacity:14} as,
\beq
\rho = \min \Big\{ \alpha\log_2\Big( 1 + 10^{0.1(SNR-\Delta)}\Big), \rho_{\rm max} \Big\},
\eeq
where $\alpha$ is the bandwidth utilization factor, $\Delta$ is
the loss factor (in dB) and $\rho_{\rm max}$ is the maximum spectral efficiency.
From \cite{AkdenizCapacity:14}, we get the values  $\Delta$ = 3 dB and
$\rho_{\rm max}$ = 4.8 bps/Hz. The value of $\alpha$ is taken as 0.83,
the same as that of LTE as reported in \cite{MogEtAl:07}.

The spectral efficiency for UL ACK ($\rho_{\rm ACK}$) is the minimum spectral efficiency
required to transmit 100 ACKs over one symbol. Thus,
$
 \rho_{\rm ACK} = \left(\frac{100 b_{\rm ACK}}{T_{\rm sym}W_{\rm tot}} \right) \approx \frac{1}{8}
$
is considered for the calculation of control overhead.

\paragraph{Control Message Size}
The LTE scheduling request is a trigger that notifies the BS that the user has data to transmit, and carries no
further information. For the design with fixed TTI we will use the same scheme for the SR, and to prevent
errors and mis-detection a 16-bit CRC is used with the SR. Thus, the size of a SR becomes 18 bits, with
2 bits set as priority bits. In order to provide the scheduler with a more complete information about the UE buffer
for the flexible TTI based design we propose that the SR should resemble the buffer status report (BSR) \cite{ts36321}. In our
analysis we consider the 8-bit short BSR and the 24-bit full BSR.
This with the CRC and the priority accounts for the SR to be either 26 or 42 bits long.

For simplicity, we consider the downlink and the uplink grant to be of the same size. In our analysis we assume
that the grants will be 80 bits long for the fixed TTI case and 100 bits long for the flexible TTI based design. The
values assumed are nearly double of those used in LTE as we are using higher order MIMO antennas and also have
a much wider bandwidth. Moreover, some additional bits are required for the flexible TTI based design to specify
the symbols (within a frame or a subframe) which are used for each of the transmissions.
The size specified includes an attached 16-bit CRC like that of LTE downlink control information (DCI).

Considering maximum spectral efficiency ($\rho_{\rm max}$), for transmission over a 1 GHz bandwidth
($W_{\rm tot}$) for a slot of period ($T$) of 125 $\mu$s,
the maximum number of bits that can be transmitted is equal to the number of available degrees of freedom $\rho_{\rm max}W_{\rm tot}T = 600,000$. This implies that a maximum of 600,000 bits can be transmitted over this time slot.
This accounts for 50 TCP packets, each 1500 bytes long. Hence, sending one HARQ acknowledgement
every ten such data units will need the transmission of 5 one-bit ACKs.

\bibliographystyle{IEEEtran}
\bibliography{bibl}

\end{document}